# Anchored-STFT and GNAA: An extension of STFT in conjunction with an adversarial data augmentation technique for the decoding of neural signals

Omair Ali[1,4†], Muhammad Saif-ur-Rehman[3,4†], Susanne Dyck[1], Tobias Glasmachers[2], Ioannis Iossifidis[3] and Christian Klaes[1]

[1] Faculty of Medicine, Department of Neurosurgery, University hospital Knappschaftskrankenhaus Bochum GmbH, Germany, [2]Institut für Neuroinformatik, Ruhr University Bochum, Germany, [3] Department of Computer Science, Ruhr-West University of Applied Science, Mülheim an der Ruhr, Germany; [4]Department of Electrical Engineering and Information Technology, Ruhr-University Bochum
† First author (these **two authors** contributed equally.)

## Abstract

Brain-computer interfaces (BCIs) enable communication between humans and machines by translating brain activity into control commands. Electroencephalography (EEG) signals are one of the most used brain signals in non-invasive BCI applications but are often contaminated with noise. Therefore, it is possible that meaningful patterns for classifying EEG signals are deeply hidden. State-of-the-art deep-learning algorithms are successful in learning hidden, meaningful patterns. However, the quality and the quantity of the presented inputs is pivotal. Here, we propose a novel feature extraction method called anchored Short Time Fourier Transform (anchored-STFT), which is an advanced version of STFT, as it minimizes the trade-off between temporal and spectral resolution presented by STFT. In addition, we propose a novel augmentation method, called gradient norm adversarial augmentation (GNAA). GNAA is not only an augmentation method but is also used to harness adversarial inputs in EEG data, which not only improves the classification accuracy but also enhances the robustness of the classifier. In addition, we also propose a new CNN architecture, namely Skip-Net, for the classification of EEG signals. The proposed pipeline outperforms all state-of-the-art methods and yields an average classification accuracy of 90.7 % and 89.54 % on BCI competition II dataset III and BCI competition IV dataset 2b, respectively.

# 1   Introduction

A brain computer interface (BCI) is used to translate neural signals into command signals to control an extracorporeal robotic device [1]. Henceforth, a BCI establishes an alternative pathway of communication and control between the user and an external machine. The successful translation of neural signals into command signals plays a vital role in the rehabilitation of physically disabled people [2, 3, 4, 5, 6, 7]. The first step in this process is the recording of neural signals from the areas of the brain which process the user's intent [3, 8, 9, 10, 11, 12, 13]. The neural signals are recorded either by invasive [5, 4] or non-invasive methods [8, 12, 13]. Invasive methods include implanting electrodes in the brain at the area of interest whereas most non-invasive BCI systems use EEG signals, i.e., the electrical brain activity recorded from electrodes which are placed on the scalp. In the next stage, the recorded signals are digitized and preprocessed using digital signal processors (DSPs). The preprocessed signals are then utilized to extract feature vectors, which are further fed to a decoding algorithm to map them to corresponding intended actions. The output of the decoding algorithm is then transformed into control signal to control the external device.

In this study, we focus on non-invasive BCI systems using EEG. EEG is one of the most common non-invasive ways of monitoring movement related neural signals [14]. Movement related signals from the motor cortex that are generated by imagining movements without any overt limb movement are called motor imagery (MI) [15, 16, 17]. In this study, we used EEG signals to decode and classify the MI signals into corresponding control signals. Classifying the MI-EEG signal is quite challenging due to two main reasons. Firstly, it has low signal-to-noise ratio. Secondly, it is a non-linear and non-stationary signal.

The successful classification of a MI-EEG signal into a corresponding control signal mainly depends on feature extraction techniques and machine learning algorithms. The current state-of-the-art feature extraction algorithms include common spatial pattern (CSP) [9, 12], short time Fourier transform (STFT) [15] and wavelet transform (WT) [16]. The conventional classifiers used to classify EEG signals [12, 18, 19] include linear discriminant analysis (LDA) [17], Bayesian classifiers [20] and support vector machines (SVM) [2, 21].

Deep-learning algorithms produced many state-of-the-art results in several computer vision tasks [22, 23]. Recently, deep learning has gained popularity in BCI and spike sorting studies [24], [25] and [26].

Similarly, in [27] a deep belief network (DBN) has outperformed SVM in the classification of MI-EEG tasks. In another study [28], DBN was used to detect anomalies in the EEG signals. In [29], DBN was also used to extract feature vectors for the classification algorithm. Convolution neural networks (CNNs) are also successfully used for decoding in BCI applications. In [30], CNN was employed in classification of MI-EEG signals. To model cognitive events from EEG signals, a novel multi-dimensional feature extraction technique using recurrent convolutional neural networks was proposed in [31].

Today, algorithms based on the CNN architecture are among the most successful algorithms in image recognition tasks. One reason behind this success is the translation invariance of CNNs. Therefore, in a few BCI studies, algorithms to convert EEG signal into image representation are proposed. In [15], the information about location, time, and frequency is combined using short time Fourier transform (STFT) to convert an EEG signal to an image structure. In [16], the MI-EEG signal is transformed into an image using a wavelet transform, only later to be used by CNN for the classification of the signal. In [32], a hybrid scale CNN architecture is presented for MI-EEG classification, which extracts the features from different frequency bands using multiple kernel scales. Furthermore, [33] reported the current state-of-the-art results for MI-EEG signals classification. Here, the authors presented a deep learning model, named EEG-Inception. EEG-Inception uses the inception layers for feature extraction. It uses the raw EEG signals as inputs and maps them to intended actions.

In this study, we presented a pipeline for MI-EEG classification, which out-performed all the current state-of-the-art studies on two publicly available datasets. The contributions of this study are:

1. STFT uses a fixed-length window and consequently presents a trade-off between temporal and spectral resolution which is critical for feature extraction. Henceforth, an extension of short time Fourier Transform (STFT), called **anchored-STFT**, is proposed for better feature extraction.
2. Obtaining large, labeled data sets is still a challenge in training deep learning models for BCI applications, henceforth a generative model-based data augmentation method called **Gradient Norm adversarial augmentation (GNAA)** is proposed that enhances the robustness and the classification accuracy of the classifier.
3. Since accurate predictions are critical for BCI applications, a shallow CNN-based architecture with few trainable parameters called **Skip-Net** is proposed which enhances the classification accuracy and avoids overfitting by adding a skip connection to a shallow architecture of CNN.

The proposed pipeline outperforms the current state-of-the-art studies [33], and [32] by achieving the average classification accuracy of **89.5 %** on BCI Competition IV dataset 2b.

## 2   Materials and Methods

In this study, the classification of MI-EEG signals is performed. The complete proposed pipeline of the classification process is shown as a block diagram in **Figure 1**. It consists of three modules: anchored-STFT as feature extraction, GNAA as data augmentation and Skip-Net for classification. We used two publicly available datasets (BCI competition IV dataset 2b, BCI competition II dataset III) which are mostly used as the benchmark for the comparison of the classification of MI-EEG signals. As we used publicly available datasets, the recording of the EEG signals is not included in the pipeline.

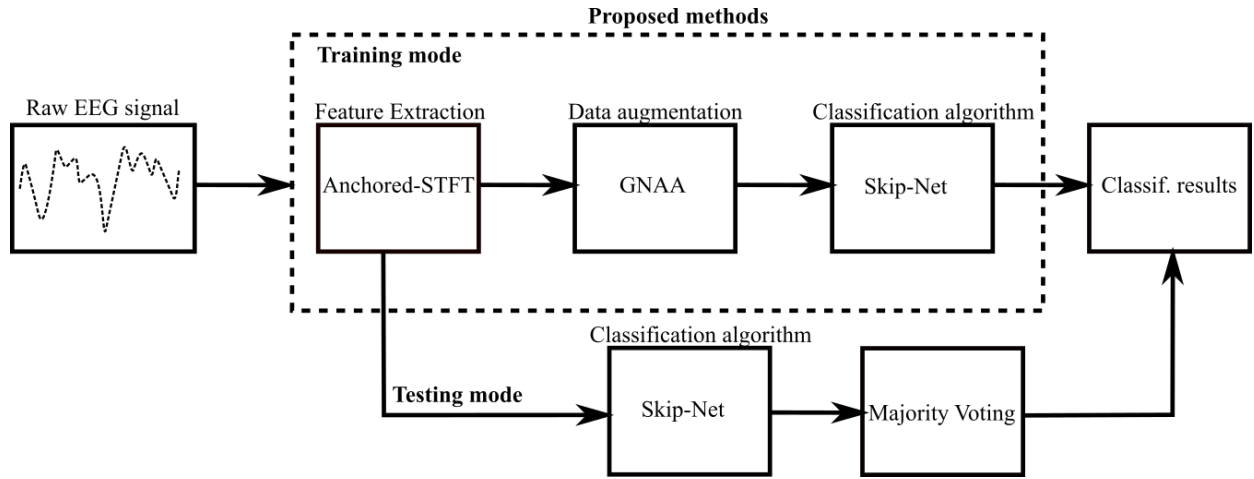

*Figure 1: The workflow of the MI-EEG signal classification process in this study. Features are extracted from raw EEG signals using anchored-STFT. During training, the GNAA method is employed on the extracted features to generate the adversarial inputs and to enhance the amount of training data to train Skip-Net algorithm. During testing, the extracted features are directly fed to the Skip-Net algorithm to perform classification and voting is done on the output of the Skip-Net algorithm to get the final classification result.*

First, the features are extracted from EEG signals using anchored-STFT. The extracted features are then used by GNAA method to generate the adversarial inputs as well as the new legitimate training examples for the Skip-Net algorithm in training mode. The extracted features from the anchored-STFT are directly used by the Skip-Net algorithm for classification during the testing mode. Voting is done on the output of the Skip-Net algorithm to get the final classification result. A detailed explanation of each of the modules of the pipeline can be found in sections below.

## 2.1 Anchored Short-Time Fourier Transform (anchored-STFT)

Short-time Fourier transform (STFT) is a variant of Fourier transform that improves the trade-off between temporal and spectral resolution. It is used for transforming non-stationary time-series signals; signals in which the frequency components vary over time, into frequency domain. STFT extracts the segments of the time-series signal by moving a window of fixed length on the time-series signal and applies the Fourier transform on each extracted segment of the signal, hence providing time-localized frequency information of the signal. On the contrary, the standard Fourier transform considers the entire signal and results in the frequency information that is averaged over the entire time domain and consequently loses the information about the time when these frequencies occurred in the time-series signal. The detailed mathematical formulation of STFT can be seen in **Supplementary Section 'Short-Time Fourier Transform (STFT)'**.

Even though STFT tries to preserve the time-localized frequency information of the signal, yet there is still a trade-off between time and frequency resolution because of a fixed-length window that transforms the time-series signal into frequency domain. The impact of the length of the window is directly proportional to frequency resolution and inversely proportional to time resolution.

As STFT uses a fixed-length window (see **Figure 2** (a 1.1)), the frequency resolution of the STFT remains same for all the locations in the spectrogram (see **Figure 2** (a 1.2)). STFT only provides a suboptimal trade-off between time and frequency resolution. Here, an extension of STFT is proposed to address this tradeoff by defining multiple anchors of variable lengths (see **Figure 2** (b)). The proposed algorithm is named as anchored-STFT. Anchored-STFT is inspired by wavelet transform [34] and Faster RCNN [23].

The working principle of anchored-STFT is as follows:

1. First, K anchors of the same shape but different lengths are defined. All the defined anchors have the same focal point (anchor position). The focal point can either be defined at the center or the left corner of the anchors (see **Figure 2** (b).
2. K is the maximum number of possible anchors, which is mathematically defined in equation (1)

$$K = \left\lfloor \frac{\log(sL)}{\log(2)} \right\rfloor \quad (1)$$

- $sL$ = length of the signal
- $aL^i$ = length of an anchor $i = 2^i$; $i=1,2, ..., K$
- Minimum length of an anchor = $minL = 2^{i=1}$
- Maximum length of an anchor = $maxL = 2^{i=K}$
- When the focal point is defined at the centre of the anchors, then the length of the anchors is given by: $aL^i$ = length of an anchor $i = 2^i + 1$; $i=1,2, ..., K$

3. The shape of the anchors could be selected by using the windows which are normally used by STFT e.g., Hann window etc.
4. N anchors are then selected from K using grid search method, where $N \subseteq K$.
5. The stride 's' by which the anchors are slid on time-series signal is half of the length of the anchor which has the smallest length among N selected anchors in case when the focal point is defined at the left corner of the anchors. In case when the focal point is at the center of the anchors, stride 's' is defined as (minL_N ± 1)/2. minL_N = minimum length of the anchor among N selected anchors. Same stride is used for all N anchors. The length of the anchors and stride determine the number of anchor positions and consequently the number of segments of time-series signal that are extracted by the anchors.
6. Zero-padding is applied to the signal to ensure that the same amount of signal segments or frames are extracted for anchors of different lengths. Zero-padding is applied either on both ends of the signal or just one end depending on whether the anchors are centered around the anchor position or cornered at the anchor position.
7. Fourier transform is applied to each segment of the time-series signal extracted by anchors and converted to frequency domain (see **Supplementary Figure 1**).
8. A separate spectrogram of the time-series signal is generated for each length anchor by aligning the spectra of adjacent, overlapping signal segments obtained by that length anchor as shown in **Supplementary Figure 1**. For example, if anchors of 4 different lengths are used, then 4 spectra of the time-series signal are generated.
9. The overlap between anchors of the adjacent anchor locations and number of anchor locations are obtained by equation (2) and equation (3) respectively.

$$overlap = aL - stride \quad (2)$$

$$no.\,of\,anchor\,locations = 1 + \frac{sL - minL\_N}{s} \qquad (3)$$

It is clear from **Figure 2** (a 1.2), that the frequency resolution of the STFT remains the same for all the locations in the spectrogram. However, it is shown in **Figure 2** (b 1.2) that an anchor (K1) of smaller length provides better time resolution and lower frequency resolution, whereas the anchor (K3) of longer length provides better frequency resolution and lower time resolution. The green and black boxes show the same frequency components computed for anchors of different lengths. Each frequency component has a different resolution for each anchor of different length which consequently provides better time-frequency resolution, which is also shown in **Figure 4**. **Figure 4** shows the input images of different time-frequency resolution generated by 5 anchors of different lengths for right-hand MI-task performed by subject 4 of BCI competition IV dataset 2b.

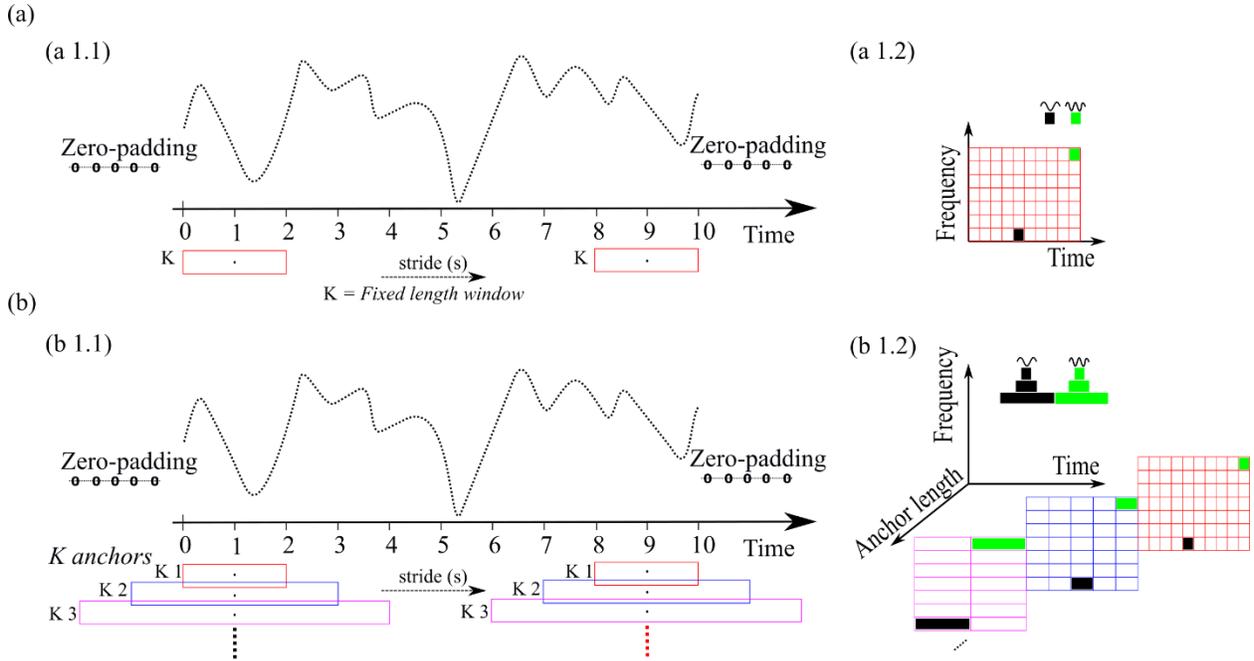

*Figure 2: Representation of time-frequency resolution of standard STFT and anchored-STFT. (a) shows the time-frequency resolution of a fixed length window K of STFT. (a 1.1) shows a fixed length window K that is convolved with the time series signal with a fixed stride (s). (a 1.2) shows the spectrogram obtained by convolving the window K with time series signal. Here, frequency resolution remains the same for all locations of the spectrogram. (b) shows the time-frequency resolution of anchored-STFT. (b 1.1) shows that anchors of different lengths are convolved with the time series signal using stride (s). (b 1.2) shows that anchor K1 with short length results into better time resolution and low frequency resolution spectrogram. Anchor K3 with longer length provides better frequency but low time resolution spectrogram. The green and black colored boxes show a frequency component computed for anchors of different lengths which in turn provides different frequency resolution for each anchor length.*

An intuitive explanation of the workflow of anchored-STFT is provided in **Supplementary Section 'Workflow of anchored-STFT'**.

## 2.2 Gradient Norm Adversarial Augmentation (GNAA)

In this study, we used the proposed GNAA method for harnessing new training inputs from the existing training inputs for the EEG data. The proposed data augmentation algorithm is different from any other existing data augmentation techniques. At first, it requires a trained neural network for the selection of meaningful features. Then, it calculates the gradient of cost function (of trained neural network) with respect to a given training input. This gradient provides the direction of the decision boundary. The given training input $x$ is slightly perturbed (by factor $\varepsilon$) towards the direction of decision boundary. As a result, it generates new inputs $x_{new}$ as shown in equation (4). 'Gradient norm' method is not only a method of generating new inputs, but it also ensures the selection of features in the given feature vector that play a pivotal role in the prediction.

$$x_{new} = x + \varepsilon \left( \frac{\frac{\partial(cost)}{\partial x}}{\left| \frac{\partial(cost)}{\partial x} \right|} \right) \quad (4)$$

We not only used equation (4) for data generation but also to study the existence of adversarial inputs in the domain of BCI studies. In this study, we define the term 'adversarial inputs' as the inputs which are modified versions of original inputs but are highly correlated. However, the employed classification algorithm fails to predict them correctly. Here, the term $\beta$ in the equation (5) defines the required minimum amount of perturbation, such that, the difference between two inputs (original input and perturbed input) remains indistinguishable in terms of correlation but the classifier can be fooled with perturbed inputs. The value of $\beta$ is (0.01) determined empirically.

$$x_{adv} = x + \beta \left( \frac{\frac{\partial(cost)}{\partial x}}{\left| \frac{\partial(cost)}{\partial x} \right|} \right) \quad (5)$$

Here, we also determine the 'pockets' of adversarial inputs. The 'pockets' are defined as the number of inputs in the train dataset that can be converted into adversarial inputs (using trained classifier) by applying the amount of perturbation defined by $\beta$ in equation (6).

Additionally, we compared the perturbation applied by the 'gradient norm' method with another existing method of crafting adversarial inputs called 'gradient sign' method [35] defined in equation (6). The perturbation applied by the two methods are significantly different as shown in **Figure 3**. The original input, applied perturbation and the new generated perturbed input by the **gradient norm method** are shown in **Figure 3** (a). Whereas the original input, applied perturbation and the new generated perturbed input by the **gradient sign method** are shown in **Figure 3** (b). The perturbation applied by the 'gradient norm' method carefully selects only features that are important for the employed classification algorithm as shown in **Figure 3** (a.2). The more important features are replaced with higher values and the value of the least important feature is slightly changed. The direction of the perturbations tends to be towards the decision boundary.

However, the perturbation applied by the 'gradient sign' method seems to be random (see **Figure 3**(b.2)). The randomness lies in the perturbation because of the signum operator in equation (6). The signum operator maps all the values greater than 0 to 1 and the values less than 0 to -1 in the perturbation matrix (see **Figure 3** (b.2)). Mathematically, the signum operator is defined in equation (7). As a result, the perturbation matrix is filled with values of either 1 or -1 and importance of each feature is disregarded.

$$x_{adv} = x + \varepsilon \, sign(\frac{\partial(cost)}{\partial x}) \tag{6}$$

$$sign := \begin{cases} -1 & if \, x < 0 \\ 0 & if \, x = 0 \\ 1 & if \, x > 0 \end{cases} \tag{7}$$

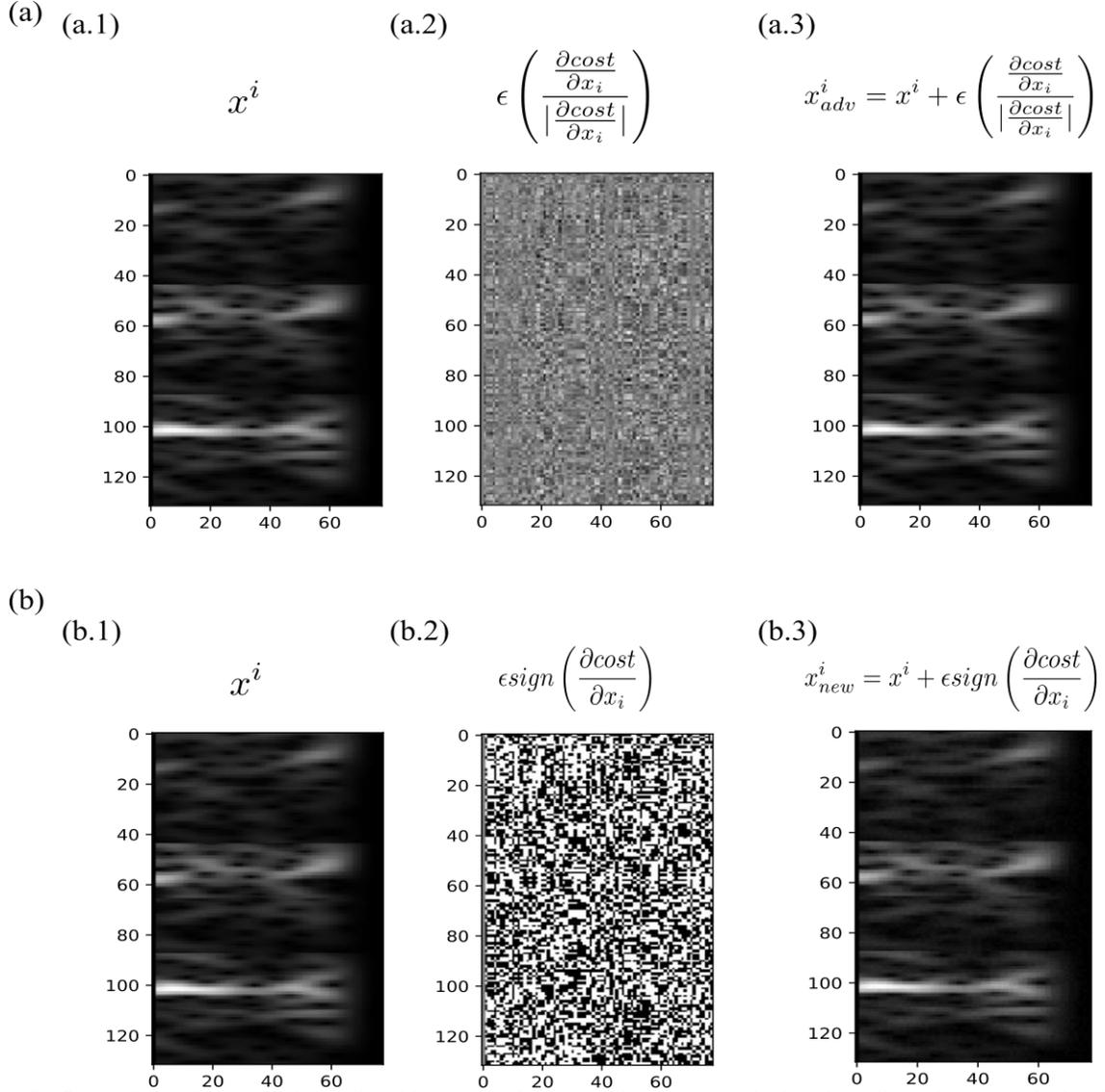

**Figure 3:** *Comparison of perturbations offered by two methods; gradient norm method and gradient signum method. (a) The original image, perturbations produced by gradient norm method and the new generated perturbed input are shown. (b) The original image, perturbations produced by gradient sign method and the new generated perturbed input are shown.*

## 2.3 Feature formation

In this study we used a convolutional neural network (CNN) based algorithm called Skip-Net for the classification of MI-EEG signals. Since the CNN based algorithms have shown state-of-art results in image recognition, therefore we also converted the EEG signals into images to use for classification by the Skip-Net algorithm.

In this study, two publicly available datasets (BCI competition IV dataset 2b [36], BCI competition II dataset III [37]) are used. The data acquisition, preprocessing and the other significant details about the datasets are discussed in detail in **Supplementary Section 'Datasets & Preprocessing'**. However, this section contains only the necessary information to extract the features from the raw EEG signals.

In case of BCI competition IV dataset 2b, the EEG signal from second 3 to second 5.5 (2.5 seconds in total) is considered for each trial and converted into frequency domain using anchored-STFT (see section 2.2). We call this interval (from second 3 to second 5.5) of the EEG signal the signal of interest (SOI) in the rest of the document. The SOI for dataset III BCI competition II lasts from second 2.75 to second 7.25. In case of 250 Hz sampling frequency, each SOI consists of 625 samples. Anchors of five different lengths are used to transform each SOI into frequency domain. So, we get five spectrums of different time-frequency resolution for each SOI. We treat these spectra as images. The lengths (in samples) of anchors used are as follows: 16, 32, 64, 128, 256. All the lengths considered are of power of 2. Stride of 8 samples is used to slide each anchor across the SOI. Here the anchors are cornered at the anchor positions. Anchor with the shortest length (8 samples) and the stride are used to determine the number of anchor positions for all the anchors and consequently the number of segments into which each SOI is divided. This results in 78 anchor locations or segments for an SOI. Since the first anchor position considered is the first sample of the SOI, so the zero-padding is only applied after the last sample of the SOI such that the 78 segments are extracted from SOI for each anchor. Equation (8) is used to calculate the zero-padding required. 257 unique FFT points as used by [15] are used to get the frequency components. This leads to a 257 x 78 image (spectrum) for each anchor, where 257 and 78 are the number of samples along the frequency and time axes, respectively.

$$Zero_{padding} = stride * (no.\ of\ anchor\ locations - 1) - signal\ length + anchor\ length \quad (8)$$

[38] has shown that mu band (8-13 Hz) and beta band (13-30 Hz) are of high interest for the classification of MI-EEG signals. Since there is an event related desynchronization (ERD) and event related synchronization (ERS) in mu and beta bands respectively when an MI task is performed, these bands are very vital for the classification of MI-EEG signals. So, we just considered these bands for further processing. Here, the mu band is represented by frequencies between 4-15 Hz and beta band is represented by the frequencies between 19-30 Hz. We then extracted the mu and beta frequency bands from each spectrum of a SOI. The size of images for extracted mu and beta frequency bands is 22 x 78 and 23 x 78, respectively. To get the equal representation of each band, we resized the beta band to 22 x 78 using cubic interpolation method. Finally, we combined these images to get an image of size $N_{fr}$ x $N_t$ (44 x 78); where $N_{fr}$ = 44 (no. of frequency components) and $N_t$ = 78 (no. of time sample points). Since, the dataset contains the EEG signals from $N_c$ = 3 electrodes ($C_3$, $C_z$ and $C_4$), we repeat the same process for all three electrodes and combine all these images from three electrodes which results in a final image of size $N_h$ x $N_t$ (132 x 78); where $N_h$ = $N_{fr}$ x $N_c$ = 132 for one anchor. We then repeat the whole process for all five anchors and get 5 images of size 132 x 78 each for each SOI. **Figure 4** shows the input images generated by using 5 anchors for an SOI of right-hand MI-task performed by subject 4.

The decrease of energy in mu band (4 -15 Hz) and increase of energy in beta band (19 - 30Hz) in the C3 channel clearly shows the ERD and ERS effect respectively for this right-hand MI-task, which is common while performing a MI-task.

Same process is done for dataset III of BCI competition II to get the input features.

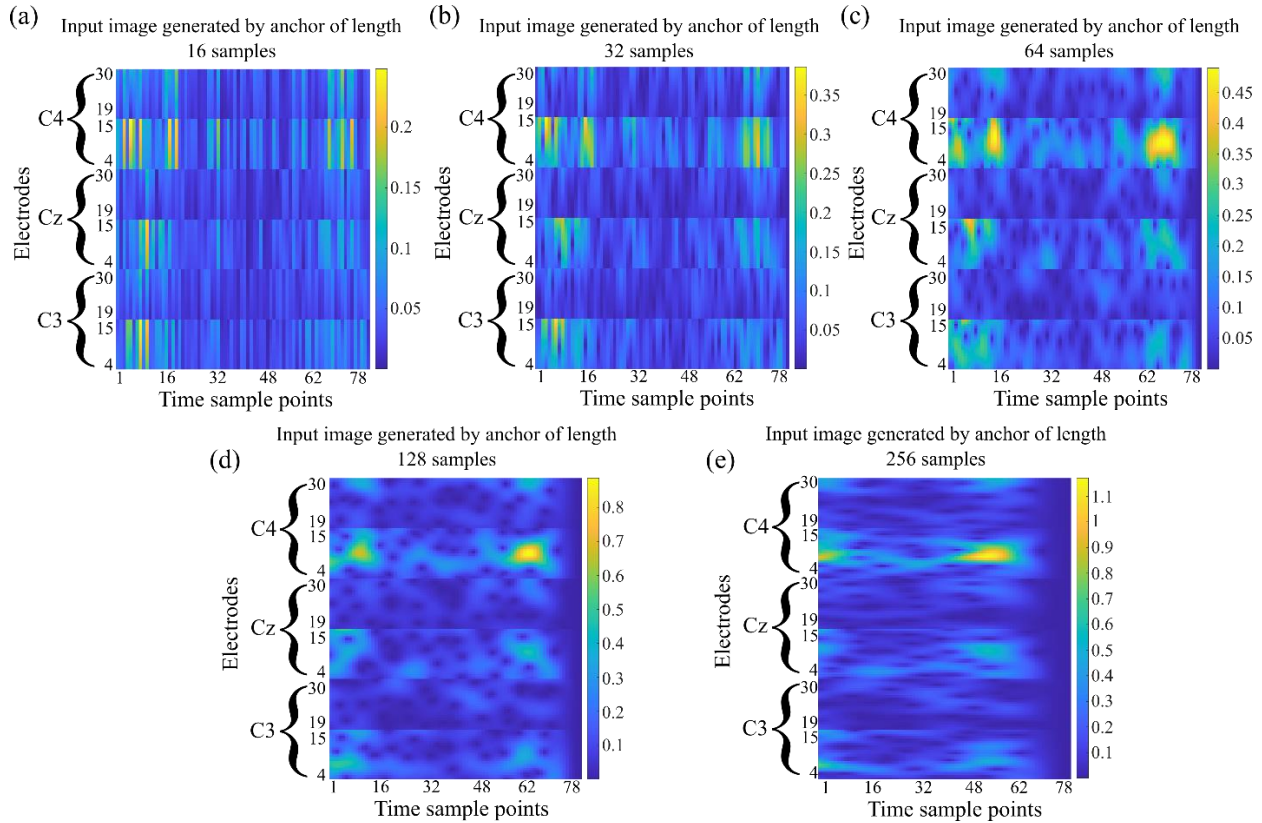

*Figure 4:* Input images generated by 5 anchors from an SOI of right-hand MI-task performed by subject 4.

## 2.4 Skip-Net

In this study, we proposed a shallow CNN-based architecture for the classification of MI-EEG signals which contains one skip connection, hence named as Skip-Net.

The Skip-Net comprises two convolutional layers. The first convolutional layer uses filters that convolve on the time axis and extracts frequency domain features along the time axis, whereas the second convolutional layer extracts the time-domain features. We used the additive skip connection to combine the extracted frequency and time domain features to prevent the loss of any information which in turn improves the classification performance of the Skip-Net compared to other classifiers. Skip-connection enhances the classification performance. The proposed architecture contains significantly less trainable parameters as compared to its counterparts proposed in [33], [32], [15] and [29]. Skip-connection as well as less parameters also reduces the risk of overfitting.

The architecture of the Skip-Net is shown in **Figure 5**. First layer in Skip-Net architecture is the input layer. The dimensions of the input layer are $N_h$ x $N_t$. The second layer is the convolutional layer which uses 16 kernels of size $N_h$x1 to convolve the input image at a stride of 1 in both horizontal and vertical directions. Rectified linear units (ReLUs) are used as the activation functions.

The output of the convolutional layer is of the size $1 \times N_t \times 16$. Batch normalization is applied at the output of the convolutional layer. The next layer is the second convolutional layer which uses 16 kernels of size 1x3 to convolve the output of the last layer in horizontal direction with a stride of 1. ReLUs are used here as the activation function and batch normalization is also applied at the output of the second convolutional layer. Next layer is the addition layer which adds the output of the first ReLU and second ReLU function. Same padding is applied in the second convolutional layer to keep the dimensions of the second convolutional feature map to be the same as the output of the first convolutional feature map so that both feature maps are compatible for the addition layer. The output of the addition layer is then fed to a fully connected layer which has 128 neurons and uses a dropout of 50 % as regularization to avoid overfitting. ReLUs are also used as activation function here. The last layer is the output layer which uses Softmax function to output the predictions. The proposed architecture is inspired by residual learning framework [39].

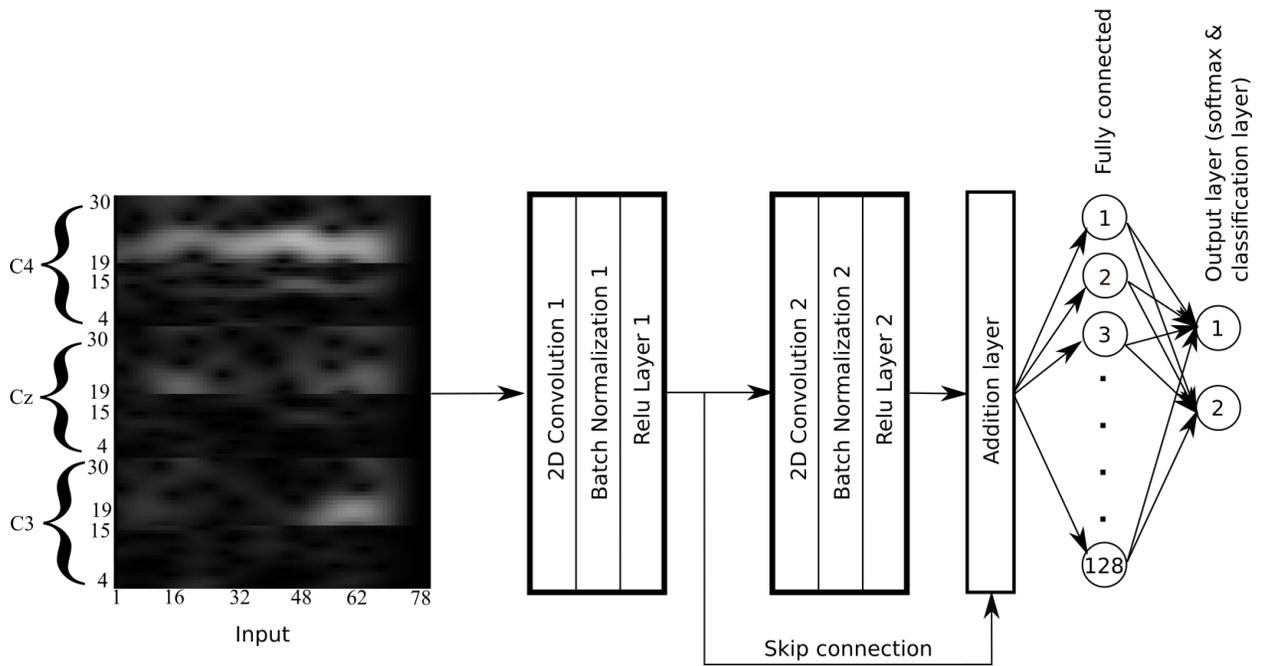

*Figure 5: Illustration of the Skip-Net architecture for the classification of MI-EEG signals.*

## 2.5 Workflow at inference time

It is shown in **Figure 1** that the features (spectra) generated by anchored-STFT are directly used by the Skip-Net algorithm to produce the classification results in test mode. As mentioned in section **Feature formation**, each SOI is transformed into 5 spectra of different time-frequency resolutions as graphically represented in **Figure 6**. Skip-Net classifies each spectrogram into one class which results in 5 predicted outputs for each SOI (one for each spectrogram). Final classification is based on majority voting using the 5 predicted outputs. The number of anchors (N) used must be odd to prevent ties. The graphical representation of the forward pass of the whole pipeline during the testing mode is shown in **Figure 6**.

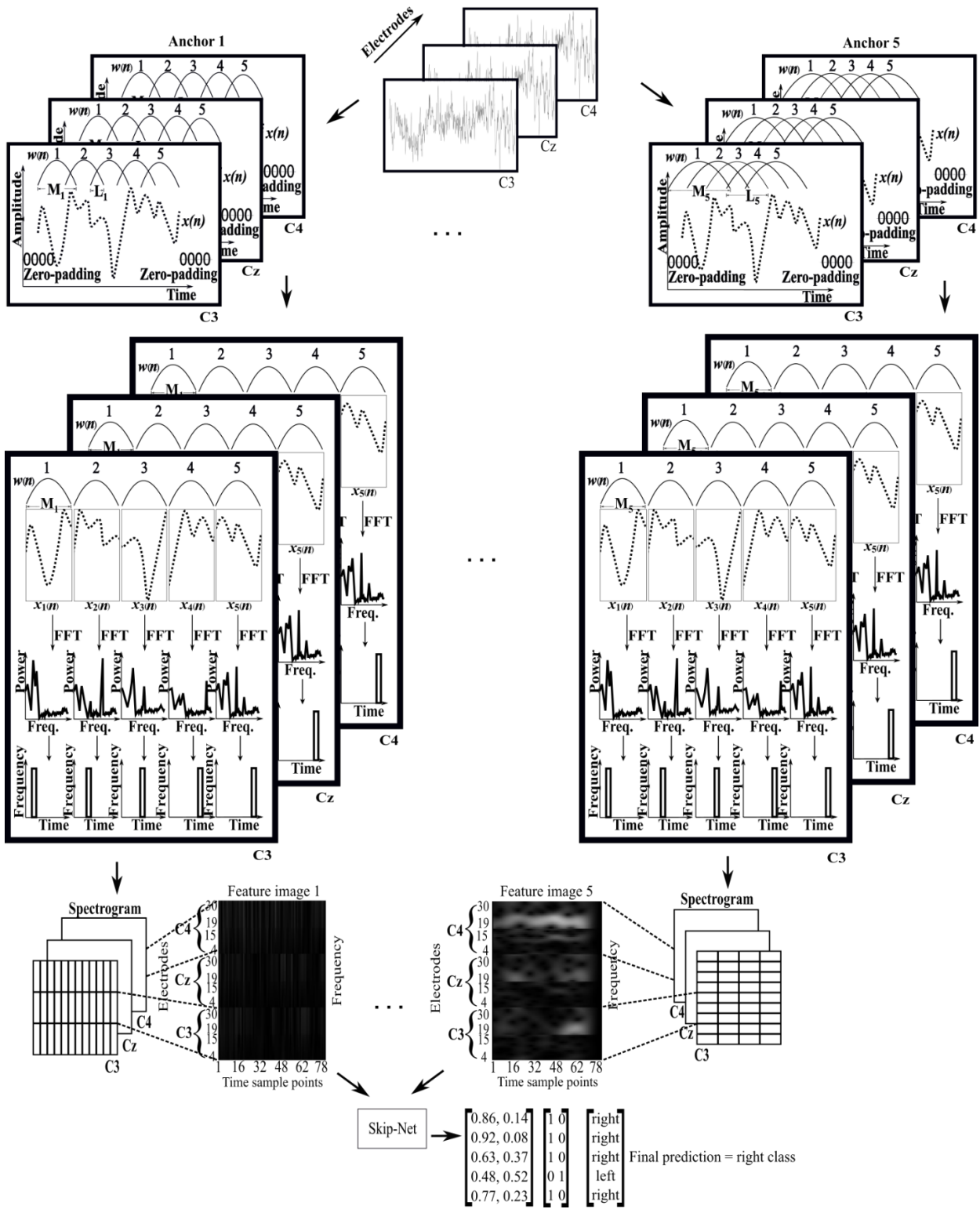

*Figure 6: Graphical representation of whole pipeline in testing mode. Five spectra are computed for each SOI for each channel. Each spectrogram is then fed to Skip-Net to make five predictions in total for each SOI. Voting is done on five output predictions. Class with maximum number of occurrences is the final predicted class for the trial.*

## 2.6 Source Code

We will upload the code and the trained models on GitHub after the successful publication of the manuscript so that others could also use it.

# 3 Results

This section includes the ablation studies to select the values of hyperparameters of proposed pipeline. In addition, this section presents the evaluation performance comparison of our proposed pipeline with other existing state-of-the-art studies.

We used two publicly available datasets (BCI competition II dataset III and BCI competition IV dataset 2b).

Most of the studies that used BCI competition IV dataset 2b for the evaluation of their methods used this dataset in two different ways. Most recent studies [33], [32] first combined the data of all the recording sessions (01T, 02T, 03T, 04E, 05E), then distributed the data into training and evaluation datasets. We named this data distribution as **data-distribution 1**. Other studies [15], [11], [40], [41], [42] used first three recording sessions (01T, 02T, 03T) for training and last two recording sessions (04E, 05E) for evaluation. We named this data distribution as **data-distribution 2.** In this study, we evaluated the proposed methods for both data distributions.

However, for dataset III of BCI competition II, all the studies used one session for training and the second session for evaluation.

## 3.1 Evaluation metrics

We used accuracy and kappa values as the evaluation metrics. Kappa value is calculated by equation (9).

$$kappa = \frac{accuracy - random\ accuracy}{1 - random\ accuracy} \quad (9)$$

## 3.2 Ablation Study

We used **data-distribution 2** of dataset 2b from BCI competition IV for the ablation studies.

### 3.2.1 Tuning of hyperparameters of anchored-STFT

Anchored-STFT includes number and combination of anchors as well as the stride as its hyperparameters. The selection of the value of hyperparameters effect the evaluation accuracy as well as the computation cost.

The total number of anchors in anchored-STFT are calculated using equation (2). The selection of number of anchors presents a trade-off between accuracy and computational cost. In principle, a greater number of anchors used results in higher classification accuracy, but it also results in higher computational cost. Increasing the number of anchors may also increase the redundancy in the extracted information which could cause the overfitting in shallow CNN architectures such as Skip-Net which in turn could decrease the overall classification accuracy. A deeper architecture with more convolutions and fully connected layers may be required to learn the hidden meaningful patterns which in turn leads to higher computational cost, that is undesirable for online decoding of neural signals in BCI applications.

To analyze the effect of different numbers and combination of anchors on the evaluation accuracy and the computation cost, several analyses are performed which investigate the relation between the numbers and combination of anchors used and their effect on the overall evaluation accuracy and the computational power. Based on the analysis presented in **Supplementary Table 1, Supplementary Table 2, Supplementary Table 3,** and **Supplementary Table 4** of **Supplementary Materials**, total number of anchors selected are 5 and the combinations used are 16,32,64,128,256.

The selection of stride is also a hyperparameter which effects the evaluation accuracy as well as the computation cost. Stride is selected based on the anchor with smallest length. The criteria for the selection of stride are such that the overlap between smallest anchor at adjacent anchor locations is 50 % minimum. However, the detail analysis of stride which results in overlap of 100 %, 75 %, 50 %, 25 % and 0 % on the overall evaluation accuracy is presented in **Supplementary Table 5** of **Supplementary Materials**. Based on the analysis, the selected stride is 8 which ensures at least the 50 % overlap between the anchor of smallest length at adjacent anchor locations. This stride ensures the optimized trade-off between the evaluation accuracy and the computation cost.

In all the remaining analyses, the values of the hyperparameters used are as such:

Anchors = [16,32,64,128,256]

Stride = 8

### 3.2.2 Performance comparison of anchored-STFT with Continuous wavelet transform (CWT) and STFT feature extraction methods and the effect of adding skip-connection to CNN architecture

Since our method is inspired from wavelet transform, and is an extension of STFT, a comparison of these methods with anchored-STFT is performed to validate the findings. **Data-distribution 2** of dataset 2b from BCI competition IV is used for this analysis. The comparison is made on two CNN based architectures i.e., proposed CNN architecture with skip connection (Skip-Net) and standard CNN architecture. Anchored-STFT using Skip-Net outperformed the STFT and CWT methods by 3.7 % and 3.6 %, respectively. However, the Anchored-STFT using standard CNN architecture outperformed the STFT and CWT methods by 3.1 % and 5.4 %, respectively. The comprehensive comparison of each subject is presented in **Supplementary Table 6**. This analysis depicts that adding a skip-connection to the standard CNN architecture yields improvement in the performance of the classifier.

### 3.2.3 Hyperparameters tuning during training for Skip-Net

The Skip-net explained in section **Skip-Net** is a deep-learning model. It involves several hyperparameters and the tuning of hyperparameters is done using grid search. The hyperparameters and their corresponding values after tuning used to train the Skip-Net algorithm are as follows:

1. Optimization algorithm = Adam
2. Momentum = 0.9
3. Initial Learning rate = 0.01
4. Learning rate drop factor = 0.5
5. Learning rate drop period = 5 epochs
6. Regularization = L2 norm (0.01), Dropout (0.5)
7. Max Epochs = 200
8. Mini batch size = 200

### 3.2.4 Evaluation of robustness of classifier using inputs generated by GNAA

It is of cardinal importance to enhance the robustness of the classifier at inference time. The data generated by GNAA improves the robustness as well as the classification accuracy. This fact is validated by a comprehensive analysis which is performed to evaluate the impact of new inputs generated by GNAA method on the robustness of the classifier. This analysis is shown in section '**Impact of inputs generated by GNAA on robustness of classifier**' of **Supplementary Materials**. In addition, a quantitative comparison of perturbations generated by GNAA, and gradient sign method is performed (see section **Impact of inputs generated by GNAA on robustness of classifier**' of **Supplementary Materials**).

The following conclusions are drawn from the aforementioned analyses:

1) The existence of adversarial inputs is not random in nature (**Figure 3** (a.2)) as produced by gradient sign method which uses the 'sign' operator (see **Figure 3** (b.2)). However, GNAA method selects only the meaningful features to perturb the inputs to generate the adversarial inputs as shown in **Figure 3** (a).
2) Training the classifier on original training data plus perturbed inputs generated by GNAA method can improve the overall average classification accuracy slightly more compared to gradient sign method, since the carefully perturbed inputs generate more training inputs that resemble closely the data distribution of the original training data.
3) Training the model on perturbed inputs along with the original training data enhances the robustness against adversarial attacks.
4) The perturbations applied by GNAA, and gradient sign method can provide the insight of the quality of the training data. As shown in **Supplementary Table 8**, subject 2 and subject 3 resulted in a greater number of adversarial examples compared to subject 4 and subject 5. It can be concluded that the discrimination power between the different classes of subject 2 and subject 3 is less as compared to subject 4 and subject 5 which is also evident from classification accuracy of these subjects as reported in **Supplementary Table 7**. It can also be inferred that, in case of subject 2 and subject 3, the feature vectors of distinct classes are quite close to the decision boundary determined by the classifier which also results in greater number of adversarial inputs when slightly perturbed.

### 3.3 Comparison of proposed pipeline with state-of-the-art studies on dataset 2b Competition IV

#### 3.3.1 Comparison with state-of-the-art studies using data-distribution 1

Here, the comparison of the proposed pipeline with most recent state-of-the-art studies reported in [33] and [32] is presented. In order to have fair comparison, we restructured the data as reported in most recent study [33]. The proposed pipeline outperformed all the state-of-the-art studies by yielding the average classification accuracy of 89.5 % which is 1 % higher than most recent results produced by EEG-inception method reported in [33], whereas it is 1.9 % higher than HS-CNN method reported in [32]. The proposed pipeline outperformed EEG-inception method for 7 out of 9 subjects whereas, it outperformed HS-CNN for 6 out of 9 subjects. In addition, the robustness is improved by employing the proposed pipeline. The enhanced robustness is depicted in the low average standard deviation among all subjects. The proposed method resulted in the standard deviation of 5.9, which is 13.8 % less than HS-CNN method reported in [32]. However, it is quite close to EEG-inception method.

In addition to improvement in the classification accuracy, the proposed method contains the lowest number of tunable parameters. The Skip-Net architecture contains 163040 tunable parameters, whereas EEG-inception model contains 25.1 % more tunable parameters. However, HS-CNN architecture contains 150.3 % more tunable parameters compared to the proposed architecture which make them more prone to overfitting and increased training and evaluation time.

*Table 1: Performance comparison of the proposed pipeline with state-of-the-art studies on dataset 2b from BCI competition IV. To have a fair analysis, we used the same data structure as reported by the current state-of-the-art study [33].*

|          | HS-CNN [32] | EEG-inception [33] | anchored-STFT + Skip-Net-GNAA |
|----------|-------------|--------------------|-------------------------------|
| S1       | 80.5        | 87.2               | **89.8**                      |
| S2       | 70.6        | 79.7               | **80.4**                      |
| S3       | 85.6        | 84.1               | 80.4                          |
| S4       | 94.6        | 96.3               | **98.2**                      |
| S5       | 98.3        | 94.0               | 94.6                          |
| S6       | 86.6        | 89.2               | **91.7**                      |
| S7       | 89.6        | 82.9               | 88.5                          |
| S8       | 95.6        | 90.6               | 90.6                          |
| S9       | 87.4        | 92.8               | 91.7                          |
| Avg.     | 87.6        | 88.5               | **89.5**                      |
| Std Dev. | 8.4         | 5.5                | 5.9                           |

#### 3.3.2 Comparison with state-of-the-art studies using data-distribution 2

Studies using data-distribution 2 for the evaluation of their methods performed two types of experiments:

1. 10-fold Cross-validation classification on training data
2. Session-to-session classification: Separate recording sessions used for training and evaluation.

### 3.3.2.1 10-fold Cross-Validation (BCI competition IV dataset 2b)

[11] introduced Filter Bank Common Spatial Pattern (FBCSP) algorithm, which is the winner algorithm of the competition, and performed 10-fold cross-validation on the training data. The cross-validation performance of proposed pipeline is compared with FBCSP algorithm and [43], which is shown **Table 2** in terms of kappa values.

Here, the average kappa value of the FBCSP method and [43] is 0.502 and 0.414 respectively, whereas the anchored-STFT + Skip-Net-GNAA obtained the average kappa value of **0.520**. The higher kappa value of the proposed methods in comparison with the other methods indicates high generalization quality. The proposed pipeline increased the kappa value by 25.6 % and 3.6 % with respect to [43] and FBCSP, respectively.

**Table 2** shows that the proposed approach outperformed FBCSP method for 6 out of 9 subjects.

*Table 2: Comparison of Kappa results generated by [43], FBCSP [11] and proposed pipeline for 10-fold cross-validation task on dataset 2b from BCI competition IV.*

| Subjects | [43] | FBCSP [11] | anchored-STFT + Skip-Net-GNAA |
|---|---|---|---|
| S1 | 0.408 | 0.546 ± 0.017 | **0.598** ± 0.074 |
| S2 | 0.224 | 0.208 ± 0.028 | 0.145 ± 0.142 |
| S3 | 0.134 | 0.244 ± 0.023 | 0.124 ± 0.163 |
| S4 | 0.778 | 0.888 ± 0.003 | **0.902** ± 0.047 |
| S5 | 0.524 | 0.692 ± 0.005 | **0.749** ± 0.055 |
| S6 | 0.414 | 0.534 ± 0.012 | **0.662** ± 0.082 |
| S7 | 0.686 | 0.409 ± 0.013 | **0.512** ± 0.060 |
| S8 | 0.236 | 0.413 ± 0.013 | **0.427** ± 0.068 |
| S9 | 0.326 | 0.583 ± 0.010 | 0.558 ± 0.073 |
| Average | 0.414 | 0.502 ± 0.014 | **0.520** ± 0.092 |

In addition to average kappa values for 10-fold cross-validation, we also compared the performance of our approach with some other methods [15], [40], [41], and [42] that provided the best kappa values for dataset 2b of BCI competition IV. We also used the best kappa values of proposed method for this comparison. Our approach outperformed the existing studies in terms of maximum kappa value comparison by yielding the maximum average kappa value of **0.737**. The detailed result of the comparison is shown in **Supplementary Table 11.**

### 3.3.2.2 Session-to-session classification performance (BCI competition IV dataset 2b)

In addition to 10-fold cross-validation, [11] also performed session-to-session classification evaluation for FBCSP and CSP algorithms. Other studies [44], [43] and [45] also performed session-to-session classification evaluation. Here, they used all the three training sessions (01T,

02T and 03T) for training and the evaluation sessions (04E and 05E) for testing their algorithm in session-to-session classification analysis.

**Table 3** shows the kappa value results of the proposed method and its comparison with methods reported in [44], [43], [45] and [11] in session-to-session classification task. Here, it is shown that proposed pipeline yielded the highest average kappa value of 0.635 compared to the other methods. It indicates that presented method provided 22.1 %, 6.0 %, 5.8 %, 61.1 % and 13.8 % improvement in terms of average kappa value with respect to CSP, FBCSP, [45], [43] and [44], respectively.

**Table 3** shows that, our method outperformed FBCSP algorithm and [45] for 6 out of 9 subjects whereas, it outperformed [44] for 8 out of 9 subjects. It outperformed the CSP algorithm and [43] for all subjects.

*Table 3: Comparison of Kappa results generated by [44], [43], CSP, FBCSP [11], and anchored-STFT + Skip-Net-GNAA for session-to-session classification task (trained on 01T, 02T and 03T sessions and evaluated on 04E and 05E sessions) of dataset 2b from BCI competition IV.*

| Subjects | [45] | CSP [11] | FBCSP [11] | [43] | [44] | anchored-STFT + Skip-Net (GNAA) |
|---|---|---|---|---|---|---|
| S1 | 0.540 | 0.319 | 0.400 | 0.406 | 0.450 | 0.500 |
| S2 | 0.290 | 0.229 | 0.207 | 0.012 | 0.128 | 0.232 |
| S3 | 0.220 | 0.125 | 0.219 | 0.056 | 0.112 | 0.194 |
| S4 | 0.930 | 0.925 | 0.950 | 0.876 | 0.944 | 0.938 |
| S5 | 0.640 | 0.525 | 0.856 | 0.276 | 0.768 | 0.844 |
| S6 | 0.690 | 0.500 | 0.613 | 0.482 | 0.574 | **0.744** |
| S7 | 0.500 | 0.544 | 0.550 | 0.238 | 0.550 | **0.638** |
| S8 | 0.820 | 0.856 | 0.850 | 0.662 | 0.838 | **0.868** |
| S9 | 0.740 | 0.656 | 0.744 | 0.544 | 0.668 | **0.756** |
| Average | 0.600 | 0.520 | 0.599 | 0.394 | 0.558 | **0.635** |

In addition, comparison is made between proposed pipeline and the algorithms presented in [15]. The proposed method outperformed all the presented algorithms in [15] including its counterparts (CNN and CNN-SAE) by providing 5.6 % and 2.9 % higher average accuracy, respectively. The detailed comparison is provided in **Supplementary Table 10**.

### 3.4 Comparison of proposed pipeline with state-of-the-art studies on BCI II, dataset III

To further validate the performance of our method, we employed our proposed pipeline on another publicly available dataset III from BCI competition II. Since this dataset is well divided into training and test data, the evaluation of the presented pipeline is trivial. Here, we only performed the evaluation on the unseen (test) dataset. The input images are computed as explained in the section **Feature formation**. **Table 4** shows the comparison of classification accuracy and kappa values on this dataset produced by proposed method, methods presented in [15] (CNN, CNN-SAE) and the winner algorithm [46] of the BCI competition II, dataset III.

**Table 4** shows that proposed method outperformed the winner algorithm and provided 1.4 % and 3.9 % improvement in terms of accuracy and kappa value, respectively. It also outperformed CNN and CNN-SAE methods by 1.4 % and 0.7 %, respectively in terms of accuracy and 3.56 % and 1.75 %, respectively in terms of kappa values.

*Table 4: Comparison of accuracy and kappa results on BCI competition II dataset III produced by anchored-STFT + Skip-Net-GNAA, CNN, CNN-SAE [15] and the winner algorithm [46].*

|  | CNN [15] | CNN-SAE [15] | winner algorithm [46] | anchored-STFT+ Skip-Net-GNAA |
|---|---|---|---|---|
| Accuracy | 89.3 | 90.0 | 89.3 | **90.7** |
| Kappa | 0.786 | 0.800 | 0.783 | **0.814** |

# 4 Discussion and summary

In order to increase the quality of neural signal we developed in the current work a novel algorithm for feature formation called anchored-STFT in conjunction with a data augmentation method named GNAA. We applied the anchored-STFT on two public available datasets which significantly improved the decoding performance of MI-EEG. We introduced a novel architecture of CNN called Skip-Net which framed the newly developed methods. We showed that the decoding accuracy on the dataset used in this study is further improved by adding the augmented data generated by GNAA in the decoding loop. Lastly, we investigated the existence of adversarial inputs in BCI applications. To the best of our knowledge, there is no other study that has investigated the existence of adversarial inputs in neural data.

The proposed anchored-STFT is inspired by wavelet transform [34] and Faster RCNN [47]. Wavelets transform scales and dilates the mother wavelet. It then slides these scaled and dilated wavelets across the time-domain signal to generate a scalogram in the frequency domain. However, anchored-STFT uses anchors of different lengths. It slides these anchors across the time-domain signal to transform it to a spectrogram with different time-frequency resolution in frequency domain. Anchored-STFT generates one spectrogram for each anchor whereas the wavelet transform produces only one scalogram for all the used scales and translation factors. The anchored-STFT also addresses the limitation of standard STFT by minimizing the trade-off between temporal and spectral resolution. Anchored-STFT uses anchors of different lengths to extract segments of corresponding lengths from the time-series signal and applies Fourier transform to each extracted segmented signal. Henceforth, temporal, and spectral resolution is optimized.

Additionally, we proposed a novel architecture for the classification of MI-EEG signals which contains one skip connection, hence named Skip-Net. Our Skip-Net comprises two convolutional layers. The first convolutional layer uses filters that convolve on the time axis and extracts frequency domain features along the time axis, whereas the second convolutional layer extracts the time-domain features. We used the additive skip connection to combine the extracted frequency and time domain features to prevent the loss of any information which in turn improved the classification performance of the Skip-Net compared to other classifiers.

The performance of deep learning algorithms is also dependent on the number of training examples. Therefore, we proposed a data augmentation technique to increase the amount of training examples. The proposed data augmentation algorithm used the objective function of the previously trained model, which is trained on the original training examples. Then, the new inputs are crafted by perturbing the original training examples towards the direction of the decision boundary of the classifier. The direction of perturbation of each new input is determined by calculating the gradient of the optimized objective with respect to its original input as defined in equation (5). The magnitude of the perturbation is kept small and defined by factor epsilon (see equation (6)).

In this study, we showed that proposed pipeline outperformed the current state-of-the-art results on two publicly available datasets as shown in **Table 1**, **Table 2**, **Table 3**, and **Table 4**. Dataset 2b from BCI competition IV and dataset III from BCI competition II are used as the benchmark for the comparison of the MI-EEG signals decoding [33], [32], [44], [15] [43] , [11], [45]. In this study, the performance comparison is made on both these datasets. However, for dataset 2b from BCI competition IV, the available studies employed **different data distributions** to validate the performance of their proposed methods. Different studies used dataset 2b from BCI competition IV differently for the comparison. [33] and [32] firstly combined all the recording sessions (training and evaluation sessions) and then randomly split them into training and evaluation datasets. However, [44], [43], [11], [45] used first three sessions (training sessions) as the training dataset and the last two sessions (evaluation sessions) as the evaluation dataset as provided and recommended by the organizers of the dataset [36]. [15] used only the training sessions for the performance validation of the proposed methods. The authors used first and second recording sessions (training sessions) for training the algorithms whereas used just the third session (third training session) for the evaluation.

In this study, we evaluated our method with the aforementioned studies using the data distribution as reported in their respective work. **Table 1** shows that the proposed method outperformed all the state-of-the-art methods such as EEG-inception and HS-CNN by obtaining 89.5 % average classification accuracy for dataset 2b from BCI competition IV. **Table 2** shows that the presented algorithm achieves 25.6 % and 3.6 % increment in cross-validation accuracy with respect to [43] and FBCSP, respectively. Similarly, **Table 3** shows that proposed pipeline yielded the highest average kappa value of 0.635 compared to the other methods. It indicates that presented method provided 22.1 %, 6.0 %, 5.8 %, 61.1 % and 13.8 % improvement in terms of average kappa value with respect to CSP, FBCSP, [45], [43] and [44], respectively. **Table 4** shows that proposed method outperformed the winner algorithm and provided 1.4 % and 3.9 % improvement in terms of accuracy and kappa value, respectively. It also outperformed CNN and CNN-SAE methods by 1.4 % and 0.7 %, respectively in terms of accuracy and 3.56 % and 1.75 %, respectively in terms of kappa values.

However, the results generated by using different data distributions for training and evaluation are fairly different as shown in **Table 1**, **Table 2**, **Table 3**, and **Table 4**. Therefore, in our perspective, using a standardized data distribution as provided and recommended by the organizers of the dataset would be more useful for fair comparison.

The current version of anchored-STFT constructs a separate feature matrix for each defined anchor and each feature matrix is provided to the classifier. Then, the voting strategy is applied to take the final decision. In the future, we are aiming to construct a single but more meaningful feature matrix from all the anchors. We believe that if all the necessary information is provided at once, it can increase the generalization quality of deep learning models. As a result, the computational cost of the proposed pipeline can also be reduced. Here, we briefly investigated the existence of adversarial inputs in neural data. However, more thorough investigation is required. Therefore, in future we are aiming to extract adversarial inputs created by different methods and try to train a more robust classifier by training it on data that has more variability.

# References

[1] B. Graimann, B. Allison and G. Pfurtscheller, Brain–Computer Interfaces: A Gentle Introduction, Berlin: Springer, 2010.

[2] A. Kübler, A. Furdea, S. Halder, E. M. Hammer, F. Nijboer and B. Kotchoubey, "A brain-computer interface controlled auditory event-related potential (p300) spelling system for locked-in patients," *Annals of the New York Academy of Sciences,* pp. doi: 10.1111/j.1749-6632.2008.04122.x. , 2009.

[3] C. Klaes, S. Kellis, T. Afalo, B. Lee, P. Kelsie, K. Shanfield, S. Hayes-Jackson, M. Aisen, C. Heck, C. Liu and R. A. Andersen, "Hand Shape Representations in the Human Posterior Parietal Cortex," *The Journal of Neuroscience,* p. 15466–15476, 2015.

[4] S. Kellis, K. Miller, K. Thomson, R. Brown, P. House and B. Greger, "Decoding spoken words using local field potentials recorded from the cortical surface," *Journal of neural engineering,* p. 056007, 2010.

[5] T. Aflalo, S. Kellis, C. Klaes, B. Lee, Y. Shi, K. Pejsa, K. Shanfield, S. Hayes-Jackson, M. Aisen, C. Heck, C. Liu and R. A Andersen, "Decoding motor imagery from the posterior parietal cortex of a tetraplegic human," *Science,* pp. 906-910, 2015.

[6] A. B. Ajiboye, F. R. Willett, D. Young, W. D. Memberg, B. A Murphy, J. P Miller, B. L Walter, J. A Sweet, H. A Hoyen, M. W Keith, P. H. Peckham, J. D Simeral and R. F. Kirsch, "Restoration of reaching and grasping movements through brain-controlled muscle stimulation in a person with tetraplegia: a proof-of-concept demonstration," *The Lancet,* vol. 389, no. 10081, pp. 1821-1830, 2017.

[7] J. Choi, S. Kim, R. Ryu, S. Kim and J. Sohn, "Implantable Neural Probes for Brain-Machine Interfaces - Current Developments and Future Prospects.," *Experimental Neurobiology,* vol. 27, no. 6, pp. 453-471, 2018.

[8] G. Pfurtscheller and F. Lopes da Silva, "Event-related EEG/MEG synchronization and desynchronization: basic principles," *Clinical Neurophysiology,* pp. 1842-1857, 1999.

[9] J. Müller-Gerking, G. Pfurtscheller and H. Flyvbjerg, "Designing optimal spatial filters for single-trial EEG classification in a movement task," *Clinical Neurophysiology,* pp. 787-798, 1999.

[10] M. Grosse-Wentrup and M. Buss, "Multiclass Common Spatial Patterns and Information Theoretic Feature Extraction," *IEEE Transactions on Biomedical Engineering,* pp. 1991 - 2000, 2008.

[11] K. Ang, Z. Chin, C. Wang, C. Guan and H. Zhang, "Filter bank common spatial pattern algorithm on BCI competition IV Datasets 2a and 2b," *Frontier in Neuroscience,* p. https://doi.org/10.3389/fnins.2012.00039, 2012.

[12] H. Ramoser, J. Muller-Gerking and G. Pfurtscheller, "Optimal spatial filtering of single trial EEG during imagined hand movement," *EEE Transactions on Rehabilitation Engineering,* p. DOI: 10.1109/86.895946, 2000.

[13] E. A. Mousavi, J. J. Maller, P. B. Fitzgerald and B. J. Lithgow, "Wavelet Common Spatial Pattern in asynchronous offline brain computer interfaces," *Biomedical Signal Processing and Control,* pp. 121-128, 2011.

[14] L. F. Nicolas-Alonso and J. Gomez-Gil, "Brain Computer Interfaces, a Review," *Sensors,* pp. https://doi.org/10.3390/s120201211-, 2012.

[15] Y. R. Tabar and U. Halici, "A novel deep learning approach for classification of EEG motor imagery signals," *Journal of Neural Engineering,* pp. DOI: 10.1088/1741-2560/14/1/016003, 2017.

[16] F. Li , F. He, F. Wang, D. Zhang, Y. Xia and X. Li, "A Novel Simplified Convolutional Neural Network Classification Algorithm of Motor Imagery EEG Signals Based on Deep Learning," *Applied Sciences,* 2020.

[17] K. Fukunaga, Introduction to Statistical Pattern Recognition, Elsevier, 2013.

[18] N. Firat Ince, S. Arica and A. Tewfik, "Classification of single trial motor imagery EEG recordings with subject adapted non-dyadic arbitrary time-frequency tilings," *Journal of Neural Engineering,* pp. doi: 10.1088/1741-2560/3/3/006., 2006.

[19] A. Schlögl, F. Lee, H. Bischof and G. Pfurtscheller, "Characterization of four-class motor imagery EEG data for the BCI-competition 2005," *Journal of Neural Engineering,* pp. https://doi.org/10.1088/1741-2560/2/4/L02, 2005.

[20] T. D. Nielsen and F. V. Jensen, Bayesian Networks and Decision Graphs, Springer-Verlag New York, 2001.

[21] C. Cortes and V. Vapnik, "Support-vector networks," *Machine Learning,* p. 273–297, 1995.

[22] Z. H. Shah, M. Müller, T.-C. Wang, P. M. Scheidig, A. Schneider, M. Schüttpelz, T. Huser and W. Schenck, "Deep-learning based denoising and reconstruction of super-resolution structured illumination microscopy images," *bioRxiv,* 2020.

[23] S. Ren, K. He, R. Girshick and J. Sun, "Faster R-CNN: Towards Real-Time Object Detection with Region Proposal Networks," *EEE Transactions on Pattern Analysis and Machine Intelligence,* vol. 39, no. 6, pp. 1137-1149, 2017.

[24] M. Saif-ur-Rehman, R. Lienkämper, Y. Parpaley, J. Wellmer, C. Liu, B. Lee, S. Kellis, R. Andersen, I. Iossifidis, T. Glasmachers and C. Klaes, "SpikeDeeptector: a deep-learning based method for detection of neural spiking activity," *Journal of Neural Engineering,* vol. 16 5, 2019.

[25] M. Saif-ur-Rehman, O. D. S. Ali, R. Lienkämper, M. Metzler, Y. Parpaley, J. Wellmer, C. Liu, B. Lee, S. Kellis, I. Iossifidis, T. Glasmachers and C. Klaes, "SpikeDeep-Classifier: A deep-learning based fully automatic offline spike sorting algorithm," *Journal of Neural Engineering,* pp. https://doi.org/10.1088/1741-2552/abc8d4, 2020.

[26] D. Issar, R. C. Williamson, S. B. Khanna and M. A. Smith, "A neural network for online spike classification that improves decoding accuracy," *Journal of Neurophysiology,* vol. 123, no. 4, pp. 1472-1485, 2020.

[27] X. An, D. Kuang, X. Guo, Zhao and L. He, "A Deep Learning Method for Classification of EEG Data Based on Motor Imagery," *Intelligent Computing in Bioinformatics,* pp. https://doi.org/10.1007/978-3-319-09330-7_25, 2014.

[28] D. F. Wulsin, J. R. Gupta, R. Mani, J. A. Blanco and B. Litt, "Modeling electroencephalography waveforms with semi-supervised deep belief nets: fast classification and anomaly measurement," *Journal of Neural Engineering,* pp. doi: 10.1088/1741-2560/8/3/036015., 2011.

[29] Y. Ren and Y. Wu, "Convolutional deep belief networks for feature extraction of EEG signal," in *International Joint Conference on Neural Networks (IJCNN)*, Beijing, 2014.

[30] H. Yang, S. Sakhavi, K. K. Ang and C. Guan, "On the use of convolutional neural networks and augmented CSP features for multi-class motor imagery of EEG signals classification," in *Annual International Conference of the IEEE Engineering in Medicine and Biology Society (EMBC)*, Milan, 2015.

[31] P. Bashivan, I. Rish, M. Yeasin and N. Codella, "Learning Representations from EEG with Deep Recurrent-Convolutional Neural Networks," *arXiv,* p. https://arxiv.org/abs/1511.06448, 2015.

[32] G. Dai, J. Zhou, J. Huang and N. Wang, "HS-CNN: a CNN with hybrid convolution scale for EEG motor imagery classification," *Journal of Neural Engineering,* vol. 17, no. 016025, 2020.

[33] C. Zhang, Y.-K. Kim and A. Eskandarian, "EEG-inception: an accurate and robust end-to-end neural network for EEG-based motor imagery classification," *Journal of Neural Engineering,* vol. 18, no. 4, 2021.

[34] L. DebnathJean and J.-P. Antoine, Wavelet Transforms and Their Applications, Louvain-la-Neuve: Physics Today, 2003.

[35] I. J. Goodfellow, J. Shlens and C. Szegedy, "Explaining and Harnessing Adversarial Examples," *arXiv,* p. arXiv:1412.6572, 2014.

[36] R. Leeb, F. Lee, C. Keinrath, R. Scherer, H. Bischof and G. Pfurtscheller, "Brain-computer communication: motivation, aim, and impact of exploring a virtual apartment," *IEEE Trans. Neural Syst. Rehabil. Eng.,* p. 15(4):473–82, 2007.

[37] A. Schlögl, "Outcome of the BCI-competition 2003 on the Graz data set," Graz University of Technology, Berlin, 2003.

[38] G. Pfurtscheller and FH Lopes Da Silva, "Event-related EEG/MEG synchronization and desynchronization: basic principles," *Clinical neurophysiology 110.11,* pp. 1842-1857, 1999.

[39] K. He, X. Zhang, S. Ren and J. Sun, "Deep Residual Learning for Image Recognition," in *IEEE Conference on Computer Vision and Pattern Recognition (CVPR)*, 2016.

[40] H.-I. Suk and L. Seong-Whan, "Data-driven frequency bands selection in EEG-based brain-computer interface," *International Workshop on Pattern Recognition in NeuroImaging. IEEE, 2011,* pp. 25-28, 2011.

[41] V. Gandhi, V. Arora, L. Behera, G. Prasad, D. Coyle and T. McGinnity, "EEG denoising with a recurrent quantum neural network for a brain-computer interface," in *In The 2011 International Joint Conference on Neural Networks. IEEE.,* 2011.

[42] S. Shahid, R. Sinha and G. Prasad, "A bispectrum approach to feature extraction for a motor imagery based brain-computer interfacing system," in *18th European Signal Processing Conference. IEEE, 2010,* 2010.

[43] H. Raza, H. Cecotti and Y. Li, "Adaptive learning with covariate shift-detection for motor imagery-based brain–computer interface," *Soft Computing,* 2016.

[44] Q. Zheng, F. Zhu and P.-A. Heng, "Robust Support Matrix Machine for Single Trial EEG Classification," *IEEE Transactions on Neural Systems and Rehabilitation Engineering,* 2018.

[45] S. Shahid and G. Prasad1, "Bispectrum-based feature extraction technique for devising a practical brain–computer interface," *Journal of Neural Engineering,* 2011.

[46] S. Lemm, C. Schäfer and G. Curio, " BCI competition 2003-data set III: probabilistic modeling of sensorimotor μ rhythms for classification of imaginary hand movements," *IEEE Trans. Biomed. Eng. 51,* pp. 1077- 80, 2004.

[47] S. Ren, K. He, R. Girshick and J. Sun, "Faster R-CNN: Towards Real-Time Object Detection with Region Proposal Networks," *arXiv,* p. arXiv:1506.01497, 2015.

[48] Y. Ren and Y. Wu, "Convolutional deep belief networks for feature extraction of EEG signal," in *International Joint Conference on Neural Networks (IJCNN)*, Beijing, 2014.

[49] S. Jirayucharoensak, S. Pan-Ngum and P. Israsena, "EEG-Based Emotion Recognition Using Deep Learning Network with Principal Component Based Covariate Shift Adaptation," *The Scientific World Journal,* p. https://doi.org/10.1155/2014/627892, 2014.

[50] E. Sejdić, I. Djurović and J. Jiang, "Time–frequency feature representation using energy concentration: An overview of recent," *Digital Signal Processing,* p. https://doi.org/10.1016/j.dsp.2007.12.004, 2009.

[51] C. Szegedy, W. Zaremba, I. Sutskever, J. Bruna, D. Erhan, I. Goodfellow and R. Fergus, "Intriguing properties of neural networks," *arXiv,* p. arXiv:1312.6199, 2014.

[52] A. Schlögl, "Outcome of the BCI-competition 2003 on the Graz data set," 2003. [Online]. Available: http://bbci.de/competition/ii/results/TR_BCI2003_III.pdf.

[53] J. B. Allen and R. R. Lawrence , "A unified approach to short-time Fourier analysis and synthesis," in *Proceedings of the IEEE 65.11*, 1977.

[54] A. Schlögl, D. Flotzinger and G. Pfurtscheller, "Adaptive autoregressive modeling used for single-trial EEG classification," *Biomedizinische Technik/Biomedical Engineering 42.6,* pp. 162-167, 1997.

[55] M.-a. Li, W. Zhu, H.-n. Liu and J.-f. Yang, "Adaptive Feature Extraction of Motor Imagery EEG with Optimal Wavelet Packets and SE-Isomap," *MDPI-Applied Sciences,* 2017.

[56] P. Gaur, R. B. Pachori, H. Wang and G. Prasad, "An empirical mode decomposition based filtering method for classification of motor-imagery EEG signals for enhancing brain-computer interface," *International Joint Conference on Neural Networks (IJCNN),* 2015.

[57] N. Lu, T. Li, X. Ren and H. Miao, "A Deep Learning Scheme for Motor Imagery Classification based on Restricted Boltzmann Machines," *IEEE Transactions on Neural Systems and Rehabilitation Engineering,* 2017.

## Acknowledgement

This work is supported by the Ministry of Economics, Innovation, Digitization and Energy of the State of North Rhine-Westphalia and the European Union, grants GE-2-2-023A (REXO) and IT-2-2-023 (VAFES).

# Figure Legends

**Figure 1: The workflow of the MI-EEG signal classification process in this study**. Features are extracted from raw EEG signals using anchored-STFT. During training, the GNAA method is employed on the extracted features to generate the adversarial inputs and to enhance the amount of training data to train Skip-Net algorithm. During testing, the extracted features are directly fed to the Skip-Net algorithm to perform classification and voting is done on the output of the Skip-Net algorithm to get the final classification result.

**Figure 2: Representation of time-frequency resolution of standard STFT and anchored-STFT.** (a) shows the time-frequency resolution of a fixed length window K of STFT. (a 1.1) shows a fixed length window K that is convolved with the time series signal with a fixed stride (s). (a 1.2) shows the spectrogram obtained by convolving the window K with time series signal. Here, frequency resolution remains the same for all locations of the spectrogram. (b) shows the time-frequency resolution of anchored-STFT. (b 1.1) shows that anchors of different lengths are convolved with the time series signal using stride (s). (b 1.2) shows that anchor K1 with short length results into better time resolution and low frequency resolution spectrogram. Anchor K3 with longer length provides better frequency but low time resolution spectrogram. The green and black colored boxes show a frequency component computed for anchors of different lengths which in turn provides different frequency resolution for each anchor length.

**Figure 3: Comparison of perturbations offered by two methods; gradient norm method and gradient signum method.** (a) The original image, perturbations produced by gradient norm method and the new generated perturbed input are shown. (b) The original image, perturbations produced by gradient sign method and the new generated perturbed input are shown.

**Figure 4: Spectral representation obtained by anchored-STFT.** Input images generated by 5 anchors from an SOI of right-hand MI-task performed by subject 4.

**Figure 5: Skip-Net architecture.** Illustration of the Skip-Net architecture for the classification of MI-EEG signals.

**Figure 6: Graphical representation of whole pipeline in testing mode.** Five spectra are computed for each SOI for each channel. Each spectrogram is then fed to Skip-Net to make five predictions in total for each SOI. Voting is done on five output predictions. Class with maximum number of occurrences is the final predicted class for the trial.

# Author information

These authors contributed equally: Omair Ali and Muhammad Saif-ur-Rehman

# Affiliations

**[1]Faculty of Medicine, Department of Neurosurgery, University hospital Knappschaftskrankenhaus Bochum GmbH, Germany**

Omair Ali, Susanne Dyck & Christian Klaes

**[2]Institut für Neuroinformatik, Ruhr University Bochum, Germany**

Tobias Glasmachers

**[3]Department of Computer Science, Ruhr-West University of Applied Science, Mülheim an der Ruhr, Germany**

Muhammad Saif-ur-Rehman & Ioannis Iossifidis

**[4]Department of Electrical Engineering and Information Technology, Ruhr-University Bochum, Germany**

Omair Ali & Muhammad Saif-ur-Rehman

# Contributions

O.A. and M.S. jointly performed the analysis and wrote the main manuscript. Also, they prepared all figures included in this work. S.D. reviewed the manuscript. T.G., I.I. and C.K. are the senior authors. They supervised the entire work and the process of data analysis. They also streamlined the ideas and reviewed the manuscript.

# Corresponding authors

Correspondence to Omair Ali or Muhammad Saif-ur-Rehman or Christian Klaes

## Competing Interests Statement
The authors declare no competing interests.

## Supplementary Information
Supplementary material is provided as a separate file. It contains the ablation study performed in this study. It also summarizes some additional experiments.